\documentclass[aps,twocolumn,nofootinbib,preprintnumbers,citeautoscript]{revtex4}
\pdfoutput=1 
\usepackage{graphicx}
\usepackage{epsfig,amsmath,amsthm}
\usepackage{subfig}
\newcommand{\be}{\begin{equation}}
\newcommand{\ee}{\end{equation}}
\newcommand{\bea}{\begin{eqnarray}}
\newcommand{\eea}{\end{eqnarray}}
\def\bse{\begin{subequations}}
\def\ese{\end{subequations}}

\def\IZ{\relax\ifmmode\hbox{Z\kern-.4em Z}\else{Z\kern-.4em Z}\fi}

\renewcommand{\d}{\partial}

\renewcommand{\(}{\left(}
\renewcommand{\)}{\right)}

\newcommand{\half}{\frac{1}{2}}

\def\half{\frac{1}{2}}

\def\bi{\begin{itemize}} \def\ei{\end{itemize}}

\def\({\left(} \def\){\right)}
\def\[{\left[} \def\]{\right]}
\def\be{\begin{equation}}
\def\ee{\end{equation}}
\def\ben{\begin{equation*}}
\def\een{\end{equation*}}
\def\beqa{\begin{eqnarray}}
\def\eeqa{\end{eqnarray}}

\begin{document}

\title{Holographic Stripes}

\author{Moshe Rozali, Darren Smyth, Evgeny Sorkin and Jared B. Stang \\
{\it Department of Physics and Astronomy, \\
University of British Columbia,\\
Vancouver, BC V6T 1Z1, Canada} }

\begin{abstract}{We construct inhomogeneous
asymptotically AdS black hole solutions corresponding to
    the spontaneous breaking 
of translational invariance and the formation of striped order in the boundary field theory.
We find that the system undergoes a second order phase transition in
 both the fixed density and fixed chemical potential ensembles, for sufficiently large values of the axion coupling. We investigate the phase structure as function of the temperature, axion coupling and the stripe width.
The bulk solutions have striking geometrical features
related to a magnetoelectric effect
 associated with the existence of a near horizon topological insulator. At low temperatures the horizon becomes highly
 inhomogeneous and tends to pinch off. }
\end{abstract}

\maketitle

\section{Introduction and Summary}
The gauge-gravity duality describes phenomena in strongly coupled field theories via their relation to classical or semi-classical gravitational systems.  From the perspective of the boundary quantum field theory, this relation can be used to
construct and study models for ill-understood phenomena which
arise in such strongly coupled systems. On the other hand, the relation to local quantum field theory helps motivate and interpret new results in classical and quantum gravity.
In this Letter  we apply holography to study the spontaneous breaking
of translation invariance and the formation of striped order.

Stripes are known to form in a variety of strongly coupled systems,
from large $N$ QCD \cite{Deryagin:1992rw, Shuster:1999tn} to systems of strongly
correlated electrons (for a review see \cite{vojta}). The 
formation of
stripes and the associated reduced dimensionality are speculated to be
related to the mechanism of
superconductivity in the Cuprates
\cite{concepts}. It is therefore useful to study striped
 phases in the holographic context.
 
Besides its interest in the boundary theory, this study has an
intrinsic interest in the bulk gravitational context\footnote{Our model
describes a black hole whose instability
to the formation of inhomogeneous structures resembles
the black string instability \cite{GL} which is known to be of the
second order for high enough dimensions \cite{CritDim}.}.
We describe striking bulk and boundary properties of
our bulk solutions, 
including frame dragging effects, the magnetic field, the curvature and the geometry.  Some of the features can be understood as the
emergence of a 
near horizon region which acts as a bulk topological insulator. The
magnetoelectric effect is 
then responsible for the patterns we observe for the bulk magnetic
field and vorticity.

Our study is facilitated by a numerical solution of the set of coupled
non-linear Einstein and matter equations in the bulk, which exhibit a
normalizable inhomogeneous mode. Previous studies of inhomogeneous
solutions in asymptotically AdS spacetimes 
concentrated on non-normalizable modes \cite{Flauger:2010tv} (i.e. {\it explicit} rather
than spontaneous breaking of translation invariance) or the study of
co-homogeneity one solutions \cite{Nakamura:2009tf, Donos:2012gg,Donos:2012wi,Iizuka:2012iv}, where one of the translational Killing
vectors is replaced by a helical Killing vector. More recently, such spontaneous breaking was exhibited in a probe model, which was shown to have a magnetic field induced lattice ground state \cite{Bu:2012mq}.
In contrast to the above, our solutions are co-homogeneity two, they backreact on the geometry,  and 
exhibit {\it spontaneous} breaking of translation invariance
 below a critical temperature.

These features 
are analyzed as a function of temperature. 
In particular, we find that the horizon of the black hole develops a
``neck'' and a ``bulge''
in the transverse direction which shrink with temperature, such that
the ratio of their sizes contracts as fast as $\sim
T^{\sigma}$, 
with an order $\sigma \sim 0.1$ exponent.
 Simultaneously, the proper length of the horizon in the
transverse direction grows at a rate $\sim 1/T^{0.1}$. However, the
curvature remains finite, and its maximal value, occurring at the
bulge, tends to a constant in the limit $T\rightarrow 0$.  

The bulk black hole solutions give rise to the holographic stripes on the
boundary, characterized by non vanishing momentum and electric
current and modulations in charge and mass density. Starting small
near $T_c$, the amplitudes of the modulations grow steadily at lower
temperatures, approaching finite values at $T\rightarrow 0$.

Finally, we study the thermodynamics of the system by constructing
phase diagrams in various ensembles.
For small values of the axion coupling, where the thermodynamic potentials in
both phases are nearly degenerate, our numerical method is not accurate
enough to sharply distinguish between weak first order and second order transitions \footnote{When the revised version of this Letter was nearly
  ready
  to be submitted the preprints \cite{Donos_stripes} appeared,
  investigating this model using different methods. These preprints also indicate the existence of 
  a second order phase transition for sufficiently large axion coupling.}.
However, for sufficiently large values of the axion coupling we
discover a clear second
order phase transition in the canonical (fixed charge), the grand canonical
(fixed chemical potential) and the micro-canonical ensembles. We describe both the finite system (of fixed length)
 and the infinite system, where we find that the dominant stripe width changes as function of temperature.
\section{The Holographic Setup}
\label{sec_setup}
The Lagrangian describing our coupled system is \cite{Donos:2011bh} 
\begin{eqnarray}
\mathcal{L}&=&\half R - \half\d^\mu\psi \d_\mu\psi-\frac{1}{4} F^{\mu\nu}F_{\mu\nu}- \,V(\psi) -\mathcal{L}_{int},\nonumber \\
V(\psi) &=& -6+ \half m^2 \psi^2,\nonumber \\
\mathcal{L}_{int}&=&\frac{1}{\sqrt{-g}}\frac{c_1}{16 \sqrt{3}} \,\psi \,\epsilon^{\mu\nu
  \rho \sigma}F_{\mu\nu}F_{\rho \sigma},
\label{Lgrn}
\end{eqnarray}
where $R$ is the Ricci scalar,  $F_{\mu\nu}$ is the Faraday tensor, 
$\mathcal{L}_{int}$ describes the axion coupling and $g$ is the
determinant of the metric.
We
use units in which the AdS radius $l^2=1/2$, Newton's constant
$8 \pi G_N= 1$, and $c=\hbar=1$, and choose $m^2 =-4$ and several
values of $c_1$. 

Perturbative instabilities
towards the formation of charge and current density waves were
identified in \cite{Donos:2011bh} for a range of wave numbers and
temperatures\footnote{ An interesting application of the instability in this
model has appeared very recently \cite{Donos:2012ra}.}.  We note the
 appearance of {\it axion electrodynamics}  in the bulk theory. It is
 curious that inhomogeneous instabilities (see also
 \cite{Nakamura:2009ts}) seem to involve the topology of the bulk
 fields in an essential  way, though the
 analysis performed to discover the instability is local in nature. We leave this mystery for future work.

In this Letter we investigate the end-point of the instability.  Part of the boundary data is the spatial periodicity, and we focus mostly on  the wave number with the largest critical temperature
$T_c$  \cite{Donos:2011bh}\footnote{Alternatively one can work in the
  conjugate ensemble where the tension in the 
spatial direction is fixed. 
We found that the order of the phase transition does not change in
this case, details will appear elsewhere \cite{us}.}. This state is a co-homogeneity two solution, thus we
construct the family of stationary solutions that emerge from the critical point, assuming
all the fields to be 
functions  of the radial coordinate $r$ and one spatial coordinate $x$. 

Our ansatz includes the scalar field $\psi(r,x)$, the gauge field components $A_t(r,x)$ and $A_y(r,x)$ and the metric
\bea
ds^2&=&-2r^2f(r)e^{2A(r,x)}dt^2+2r^2e^{2C(r,x)}(dy-W(r,x)dt)^2 \nonumber \\
&+&e^{2B(r,x)}\left(\frac{dr^2}{2r^2f(r)}+2r^2dx^2\right) ,
\label{ds}
\eea
where for the sake of convenience we included in the definition of the
metric functions the factor $f(r)$ characterizing 
the metric of the AdS Reissner-Nordstr\"{o}m (RN for short) solution, with horizon at $r=r_0$:
\be
f(r) = 1 - \left( 1 + \frac{\mu^2}{4r_0^2} \right) \left( \frac{r_0}{r} \right)^3 +\frac{\mu^2}{4r_0^2}\left( \frac{r_0}{r} \right)^4. \nonumber
\ee
The inhomogeneous solutions reduce to the RN solution above the critical
temperature. 

The conformal in $r,x$ plane ansatz (\ref{ds}) is convenient in
constructing co-homogeneity two solutions. With this ansatz, the Einstein
and matter equations reduce to seven coupled elliptic
equations and two constraint equations. Moreover, the constraint
system can be solved elegantly using its similarity to a
Cauchy-Riemann problem \cite{Wiseman:2002zc}. 

The boundary conditions we impose correspond to regularity conditions
at the horizon and asymptotically AdS conditions at the conformal
boundary.  
With these boundary conditions, the set of solutions we find depend on 
three parameters: the temperature $T$, the chemical potential $\mu$
and the periodicity 
in the $x$ direction $L$. Using the conformal symmetry inherent in
asymptotically AdS spaces, the moduli space of solution depends only
on the two dimensionless combinations of these parameters. To focus on the dominant critical mode
that becomes unstable at the largest temperature $T_c$ we choose $L=2\pi/k_c$.

On the spatial boundaries it  is useful to impose ``staggered" periodicity conditions. Using two reflection symmetries which are preserved by the form of the unstable perturbation, one can reduce the numerical domain to a quarter period and impose\footnote{Our boundary conditions do not exclude the homogeneous solution, but since that solution is unstable we find that in practice our numerical procedure converges to the inhomogeneous solution unless we are very close to the critical point.}
$\partial_x\psi(x=0) =0,~\psi(x={L}/{4})=0,$ 
$h(x=0) = 0,~ \partial_x h(x={L}/{4}) = 0 $ and 
$\partial_x g(x=0) = 0,~\partial_x g(x={L}/{4}) = 0$,
where $h$ represents the fields $A_y$ and $W$, and $g$ refers collectively to  $A,B,C$ and  $A_t$.

The elliptic equations derived from (\ref{Lgrn}) are discretized using finite
  difference methods and are solved numerically by a straightforward relaxation with the
specified boundary conditions.
In this method the equations are iterated starting with an initial
guess for all fields, until successive changes in the functions drop
below the desired tolerance. We verify that the remaining two constraints
are satisfied by those solutions.  Full details of our procedure are
given in the upcoming \cite{us}.
\begin{figure}[t!] 
\hspace{-.5cm}
    \includegraphics[width=9.5cm]{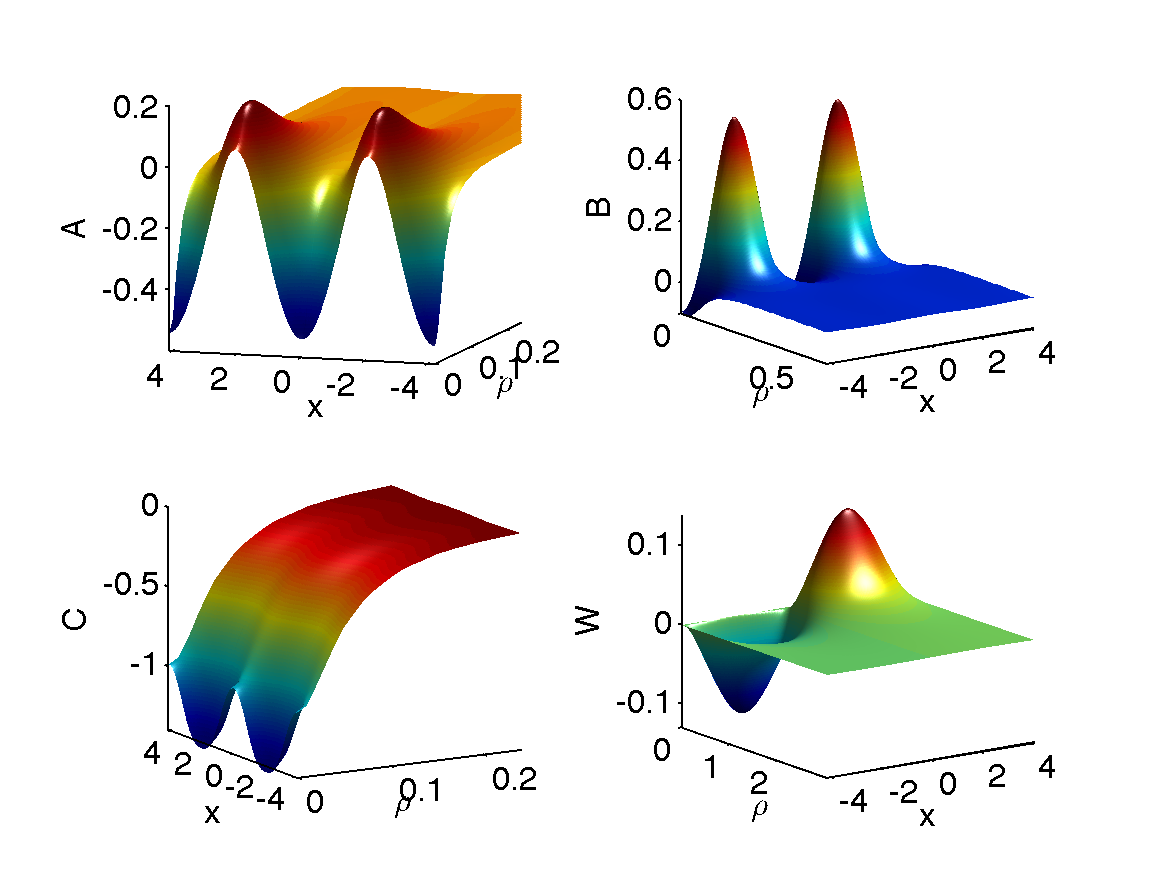}
    \caption[]{Metric functions for $\theta\simeq 0.11$ and $c_1=4.5$. Note that the
      metric functions $A, B$ and $C$ have half the period of $W$. The
    variation is maximal near the horizon, located at $\rho=0$, and it
  decays as the conformal boundary is approached, when $\rho
  \rightarrow \infty$.  The matter fields (not shown) behave in a 
  qualitatively similar manner.}
\label{fig_ds_3D}
\end{figure}
%

\section{The Solutions}
\label{sec_solutions}
A convenient way to parametrize our inhomogeneous solutions is by
the dimensionless temperature  $\theta =T/T_c$, relative to the critical
temperature $T_c$. Our method allows us to find solutions in the
range $0.003 \lesssim \theta \lesssim 0.9$ for $c_1=4.5$ and the range
$0.00016 \lesssim \theta \lesssim 0.96$ for $c_1=8$, for fixed  $\mu$.

\paragraph*{\bf{Bulk Geometry.}}  For subcritical temperatures, as we descend into the inhomogeneous regime, the
metric and the matter fields start developing increasing variation in $x$.
Fig. \ref{fig_ds_3D} displays the metric functions
for $\theta\simeq
0.11$, over a full period in the $x$ direction, in the case $c_1=4.5$. The matter fields have
qualitatively similar behaviour. The variation of all
fields is maximal near the horizon of the black hole at $\rho\equiv \sqrt{r^2-r_0^2}=0$, and
it gradually decreases toward the conformal boundary, $\rho
\rightarrow \infty$.

Many of the special features of the solutions we find are related to
the presence of axion electrodynamics, the effective description of
the electromagnetic response of a topological insulator, in the
gravity action. In the broken phase we have an axion gradient in
the near horizon geometry, which therefore realizes a topological
insulator interface\footnote{It would be interesting to discuss
  localized matter excitations on the interface, especially fermions,
  along the lines of \cite{me}. 
 }. The presence and the pattern of a near horizon magnetic field,
 summarized in the
 field $A_y$,  can be related to the {\it magnetoelectric} effect in such interfaces.

In curved space the magnetic field is accompanied by  {\it vorticity},
which is manifested by the function $W$. 
This  causes frame dragging effects in the $y$
direction. Test particles will be pushed along $y$ with speeds
$W(r,x)$, in particular the direction of the flow reverses every half the
period along $x$.   The drag vanishes at the horizon and at
the location of the nodes of $W$ where $x=L n/2$, for integer $n$
(see Fig. \ref{fig_ds_3D}).  In general,  the dragging effect remains
bounded, the vector $\d_t$ is everywhere
 timelike, and no ergoregion forms.

\begin{figure}[t!] 
    \includegraphics[width=8cm]{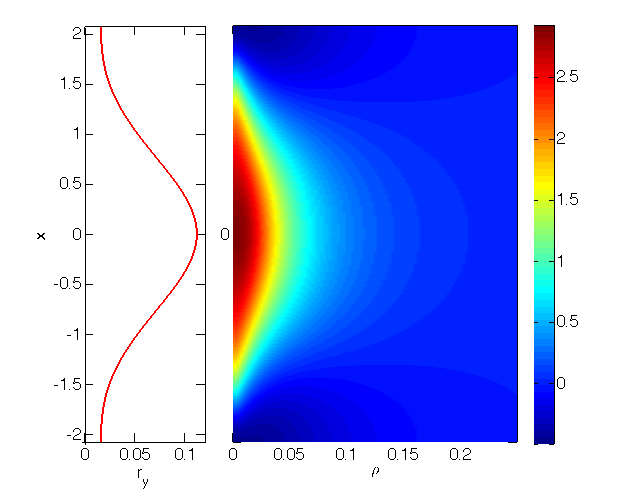}
    \caption{Left panel: The variation along $x$ of the size of the horizon
      in the $y$ direction 
includes alternating ``necks'' and ``bulges''.
  Right panel: Ricci scalar relative to that of RN black hole, $R/R_{RN}-1$ for $\theta\simeq 0.003$ over half the period.
      The scalar curvature is maximal along the horizon at the bulge $x=n\,
      L/2$ for integer $n$.  The axion coupling here is
      $c_1=4.5$ and  similar results appear for other $c_1$'s.}
\label{fig_Ricci}
\end{figure}

 The Ricci scalar of the RN solution is $R_{RN}=-24$, constant in $r$ and
  independent of the parameters of the
  black hole. This is no longer true for the inhomogeneous phases, where
  the Ricci scalar becomes position dependent. The right panel of Fig. 
  \ref{fig_Ricci} illustrates the spatial variation of the Ricci
  scalar, relative to the $R_{RN}$ for $\theta\simeq 0.003$. 
The plot corresponds to $c_1=4.5$, however we observe qualitatively
similar results for other values of the coupling. 

The
  maximal curvature is always along the horizon at  $x=n\, L/2$ for
  integer $n$. It grows when the temperature decreases and approaches
  the finite value of $R \simeq -94$ in the small temperature limit.  

The left panel in Fig. \ref{fig_Ricci}  shows the variation of transverse
extent of the horizon in the $y$ direction, $r_y(x) \equiv
\sqrt{2}\, r_0\, \exp[C(r_0,x)] $, along $x$ for
$\theta\simeq0.003$. 
 Typically there is a ``bulge''  occurring  at $x=
n\, L/2$ and a ``neck'' at  $x= (2\,n+1)\, L/4$, for integer $n$.
Note that Ricci scalar curvature is maximal at the bulge and not at the neck
as would happen, for instance, in the spherically symmetric black string case.
The size of both the neck and the bulge monotonically decrease with
temperature, however, the neck is shrinking faster. 
We find that the ratio scales as a power law $r_y^{\rm neck}/r_y^{\rm bulge} \sim
\theta^{\sigma}$ 
near the lower end of the range of $\theta$'s that we investigated. 
The exponent $\sigma$ depends on the coupling, ranging from about
$0.5$ for $c_1=4.5$ to approximately $0.1$ for $c_1=8$.

Another aspect of the geometry is the proper size of the stripe in the
$x$ direction at fixed $r$, $ l_x(r) \equiv \int_0^L
\exp[B(r,x)]\,dx.$ The proper length tends to the
coordinate length as $1/r^3$ asymptotically as $r \rightarrow \infty$,
but it exceeds that as the horizon is approached. Namely, the inhomogeneous phase ``pushes space'' around it
along $x$, resembling the ``Archimedes effect''. The proper length
of the horizon is maximal and it grows as the temperature
decreases.  We find that at small $\theta$
the proper length of the horizon diverges approximately
as $\sim \theta^{-0.1}$. 

\paragraph*{\bf{Boundary Observables.}}
Near the conformal boundary the fields decay to their AdS values, and
the subleading terms in their variation are used to define the
asymptotic charge densities of our solutions. The subleading fall-offs of the metric functions in our ansatz determine the 
boundary stress-energy tensor, whereas the fall-offs of the gauge field determine the charge and current 
densities of the boundary theory. Finally, the subleading term of the scalar field near infinity determines the scalar condensate.

For our inhomogeneous solutions we find that all charge and current densities  are spatially modulated, except for $\langle T_{xx} \rangle$, which is constant,
consistent with the conservation of boundary energy-momentum. We define the total charges of a single stripe by integrating the charge
densities over the full period $L$. These integrated quantities are
charge densities per unit length in the translationally invariant
direction $y$.

\section{Thermodynamics}
We demonstrated that below the critical temperature $T_c$ there
exists a new branch of solutions
which are spatially inhomogeneous. The question of which solution
dominates the thermodynamics depends on the ensemble used.  We start
our discussion by fixing the boundary periodicity, corresponding to working in a finite system of length $L$\footnote {We mostly discuss the case $L=\frac{2 \pi}{k_c}$, where $k_c$ is the wavelength of the dominant instability, that with the highest critical temperature.  Results for other values of $L$  will appear elsewhere \cite{us}, and are qualitatively similar.} . We  discuss the system with infinite length in the inhomogeneous  $x$-direction below.

 \begin{figure}[t!]
   \centering
    \hspace{-0pt}
    \includegraphics[width=0.46 \textwidth]{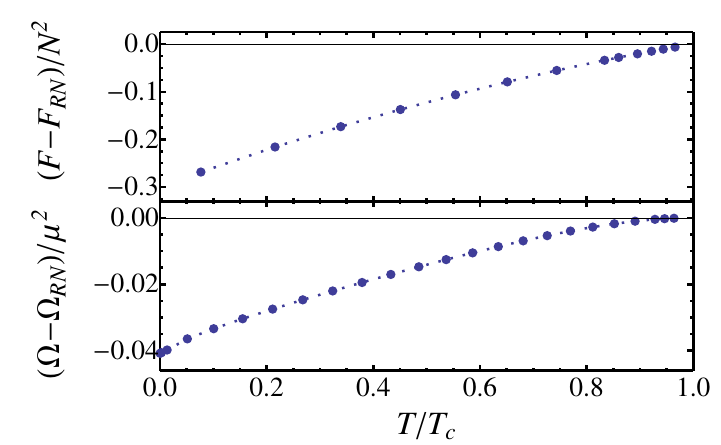}
     \vspace{-0pt}
    \caption{Difference in the thermodynamic potentials
 between the inhomogeneous phase and the RN solution for $c_1=8$, plotted against
 the temperature. In both ensembles there is a second order phase transition, with the inhomogeneous solution dominating below the critical temperature. }
    \label{free}
    \end{figure} 

 The canonical ensemble corresponds to fixing the temperature and the total charge. This
describes the physical situation in which the system is immersed in a heat bath consisting of uncharged particles.   
In the upper panel of Fig. \ref{free} we plot the difference of the normalized total free energy, $F=M-T S$, between the two classical solutions as function of the temperature
$T$, for $c_1=8$. In our ensemble the total charge $N$ is fixed, and we use the scaling symmetry of the boundary theory to set $N=1$, or in other words measure all quantities in terms of $N$. As a result the free energy is a function of one parameter, the temperature $T$. The figure displays a second order phase transition, where the inhomogeneous solution dominates the thermodynamics below the critical temperature $T_c$, the temperature at which inhomogeneities first develop.

If we fix the chemical potential instead of the charge, we discuss a
situation where the system is immersed in a plasma made of charged
particles. To study the thermodynamics we use the grand canonical
free energy $\Omega = M-TS-\mu N$,  displayed in the lower panel of
Fig. \ref{free}.
 In this ensemble it  is convenient
to measure all quantities in units of the fixed chemical potential
$\mu$. Then, again, the free energy is a function of only the
temperature $T$. In the fixed
chemical potential ensemble we find a similar second order transition, where the inhomogeneous charge distribution starts dominating the thermodynamics at the temperature
where the inhomogeneous instability develops.  

 \begin{figure}[t!]
   \centering
    \hspace{-0pt}
    \includegraphics[width=0.46 \textwidth]{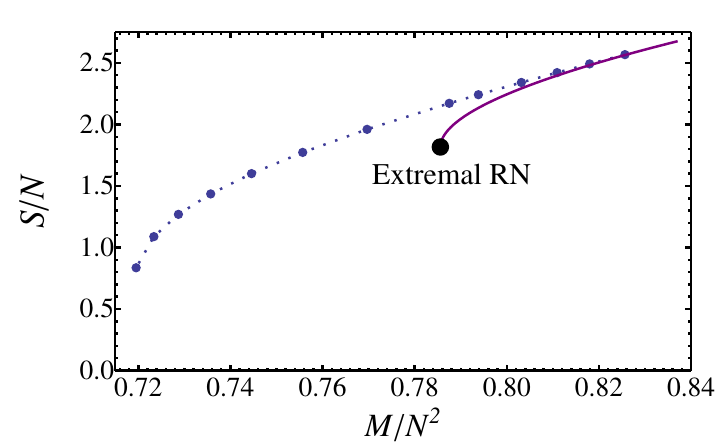}
     \vspace{-0pt}
    \caption{The entropy of the inhomogeneous solution for $c_1=8$ (points with dotted line) and of the RN solution (solid line). Below the critical temperature, the striped solution has higher entropy than the RN. The RN branch terminates at the extremal RN black hole, while the striped solution persists to smaller energies.}
    \label{entropy}
    \end{figure} 

The physical situation relevant to the study of the real time
dynamics of the instability corresponds to fixing the mass and the
charge. This is the microcanonical ensemble, describing an isolated
system in which all conserved quantities are fixed. In this ensemble
it is convenient to measure all quantities in terms of the (fixed)
charge, and the remaining control parameter is then the mass $M$. 
We find that in this ensemble as well, the striped solutions dominate the thermodynamics (have higher entropy) for all temperature below the critical temperature $T_c$,   at least when the axion coupling $c_1$ is sufficiently large. 
 This is shown in Fig. \ref{entropy}.

\begin{figure}[t!]
   \centering
    \hspace{-0pt}
    \includegraphics[width=0.42 \textwidth]{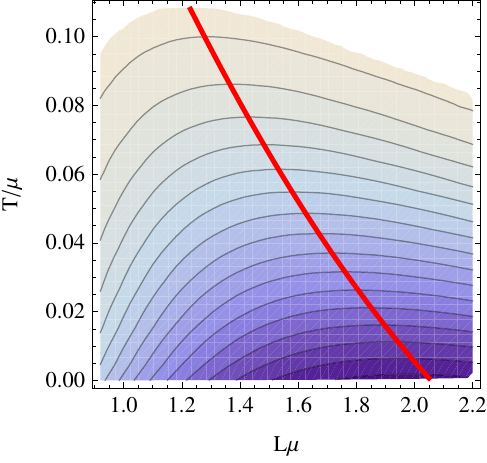}
     \vspace{-0pt}
    \caption{A contour plot of the free energy density, relative to the homogenous solution. The red line shows the variation of the dominant stripe width as function of the temperature for $c_1=8$. }
    \label{omega}
    \end{figure}

Finally, we can also study the infinite system in the inhomogeneous
$x$-direction, which we choose to look at in the canonical ensemble. In
this case we are in a position to compare the free energy {\it
  density} of different stripes, of different lengths in the $x$-direction. 
This comparison is shown in Fig. \ref{omega}, where we see that the
qualitative picture is the same as in the finite system  -- a second
order transition with striped solutions dominating at every
temperature below the critical temperature. Just below the critical
temperature, the dominant stripe is that corresponding to the critical
wavelength $k_c$. However, for lower temperature different stripes
will dominate, in fact we see in Fig. \ref{omega} that the dominant stripe width tends to increase with decreasing temperature.

\,
\,
\,
\,
\,
\,
\,
\,
\,

{\bf Acknowledgments.} We are supported by a discovery grant from NSERC of Canada. ES is
partly supported by CITA National Fellowship.  We have
benefitted from conversations with Matt Choptuik, Aristomenis Donos,  Marcel Franz,  Jerome Gauntlett, Jorge Santos,
Gordon Semenoff, Toby Wiseman and Mark van Raamsdonk.


\begin{thebibliography}{99}

\bibitem{Deryagin:1992rw} 
  D.~V.~Deryagin, D.~Y.~.Grigoriev and V.~A.~Rubakov,
  ``Standing wave ground state in high density, zero temperature QCD at large N(c),''
  Int.\ J.\ Mod.\ Phys.\ A {\bf 7}, 659 (1992).]  


\bibitem{Shuster:1999tn} 
  E.~Shuster and D.~T.~Son,
  ``On finite density QCD at large N(c),''
  Nucl.\ Phys.\ B {\bf 573}, 434 (2000)
  [hep-ph/9905448].
  
 \bibitem{vojta}
  M. Vojta,
  ``Lattice symmetry breaking in cuprate superconductors: stripes, nematics, and superconductivity,''
  Advances in Physics, Volume 58, Issue 6, 2009
  [arXiv:0901.3145].
  
  \bibitem{concepts}
  	Carlson, E. W., Emery, V. J., Kivelson, S. A., Orgad, D.,
  ``Concepts in High Temperature Superconductivity,''
  	[arXiv:cond-mat/0206217] 

\bibitem{GL} 
  R.~Gregory and R.~Laflamme,
  ``Black strings and p-branes are unstable,''
  Phys.\ Rev.\ Lett.\  {\bf 70}, 2837 (1993)


\bibitem{CritDim} 
  E.~Sorkin,
  ``A Critical dimension in the black string phase transition,''
  Phys.\ Rev.\ Lett.\  {\bf 93}, 031601 (2004)
\bibitem{Flauger:2010tv} 
  R.~Flauger, E.~Pajer, S.~Papanikolaou and ,
  ``A Striped Holographic Superconductor,''
  Phys.\ Rev.\ D {\bf 83}, 064009 (2011)
  [arXiv:1010.1775 [hep-th]].
    
\bibitem{Nakamura:2009tf} 
H.~Ooguri and C.~-S.~Park,
  ``Holographic End-Point of Spatially Modulated Phase Transition,''
  Phys.\ Rev.\ D {\bf 82}, 126001 (2010)
  [arXiv:1007.3737 [hep-th]];
  
  \bibitem{Donos:2012gg} 
  A.~Donos and J.~P.~Gauntlett,
  ``Helical superconducting black holes,''
  Phys.\ Rev.\ Lett.\  {\bf 108}, 211601 (2012)
  [arXiv:1203.0533 [hep-th]].
\bibitem{Donos:2012wi} 
  A.~Donos and J.~P.~Gauntlett,
  ``Black holes dual to helical current phases,''
  arXiv:1204.1734 [hep-th].
\bibitem{Iizuka:2012iv} 
  N.~Iizuka, S.~Kachru, N.~Kundu, P.~Narayan, N.~Sircar and S.~P.~Trivedi,
  ``Bianchi Attractors: A Classification of Extremal Black Brane Geometries,''
  JHEP {\bf 1207}, 193 (2012)
  [arXiv:1201.4861 [hep-th]].

\bibitem{Bu:2012mq} 
  Y.~-Y.~Bu, J.~Erdmenger, J.~P.~Shock and M.~Strydom,
  ``Magnetic field induced lattice ground states from holography,''
  arXiv:1210.6669 [hep-th].

\bibitem{Donos_stripes} 
  A.~Donos,
  ``Striped phases from holography,''
  arXiv:1303.7211 [hep-th];
  B.~Withers,
  ``Black branes dual to striped phases,''
  arXiv:1304.0129 [hep-th].
  
\bibitem{Donos:2011bh}
  A.~Donos and J.~P.~Gauntlett,
  ``Holographic striped phases,''
  JHEP {\bf 1108}, 140 (2011)
  [arXiv:1106.2004 [hep-th]].
  
\bibitem{Nakamura:2009ts} 
  S.~Nakamura, H.~Ooguri and C.~-S.~Park,
  ``Gravity Dual of Spatially Modulated Phase,''
  Phys.\ Rev.\ D {\bf 81}, 044018 (2010)
  [arXiv:0911.0679 [hep-th]].
  
\bibitem{Donos:2012ra} 
  A.~Donos and S.~A.~Hartnoll,
  ``Universal linear in temperature resistivity from black hole superradiance,''
  arXiv:1208.4102 [hep-th].
  
\bibitem{Wiseman:2002zc} 
  T.~Wiseman,
  ``Static axisymmetric vacuum solutions and nonuniform black strings,''
  Class.\ Quant.\ Grav.\  {\bf 20}, 1137 (2003)
  [hep-th/0209051].
  
  \bibitem{us}  
  Moshe Rozali, Darren Smyth, Evgeny Sorkin and Jared B.~Stang, to appear.
  
\bibitem{me} 
  M.~Rozali,
  ``Compressible Matter at an Holographic Interface,''
  arXiv:1210.0029 [hep-th].
  

  
    
 
  
\end{thebibliography}
\end{document}